\documentclass[aps, prd, onecolumn, 
nofootinbib,
preprint, preprintnumbers, 
superscriptaddress 
]{revtex4}

\usepackage[utf8]{inputenc}
\usepackage[T1]{fontenc}
\usepackage{lmodern}
\usepackage{microtype}
\usepackage[english]{babel}

\usepackage{amssymb,amsbsy,amsmath,amsfonts,amsthm}

\usepackage{yhmath} 

\usepackage{slashed}
\usepackage{bm}
\usepackage{times}
\usepackage{multirow}

\usepackage{graphicx}
\usepackage{epstopdf}

\usepackage{hyperref}
\hypersetup{
  colorlinks=true,
  linkcolor=blue,  
  citecolor=cyan,
  urlcolor=red,           
}

\usepackage{url}

\def\orcid#1{\kern .08em\href{https://orcid.org/#1}{\includegraphics[keepaspectratio,width=0.7em]{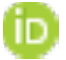}}}

\begin{document}

\title{The $\bm{\Lambda(1405)}$ in resummed chiral effective field theory}

\author{Xiu-Lei Ren 
\orcid{0000-0002-5138-7415}
}
\affiliation{Institut f\"ur Kernphysik \&  Cluster of Excellence PRISMA$^+$,
 Johannes Gutenberg-Universit\"at  Mainz,  D-55128 Mainz, Germany}

\author{E.~Epelbaum
\orcid{0000-0002-7613-0210}
}
 \affiliation{Institut f\"ur Theoretische Physik II, Ruhr-Universit\"at Bochum,  D-44780 Bochum,
 Germany}
\author{J.~Gegelia
\orcid{0000-0002-5720-9978}
}
 \affiliation{Institut f\"ur Theoretische Physik II, Ruhr-Universit\"at Bochum,  D-44780 Bochum,
 Germany}
\affiliation{Tbilisi State  University,  0186 Tbilisi,
 Georgia}
 
 \author{U.-G.~Mei\ss ner
 \orcid{0000-0003-1254-442X}
 }
 \affiliation{Helmholtz Institut f\"ur Strahlen- und Kernphysik and Bethe
   Center for Theoretical Physics, Universit\"at Bonn, D-53115 Bonn, Germany}
 \affiliation{Institute for Advanced Simulation, Institut f\"ur Kernphysik
   and J\"ulich Center for Hadron Physics, Forschungszentrum J\"ulich, D-52425 J\"ulich,
Germany}
\affiliation{Tbilisi State  University,  0186 Tbilisi,
 Georgia}

\begin{abstract}
We study the unitarized meson-baryon scattering  amplitude at leading order in the strangeness $S=-1$
sector using  time-ordered perturbation theory for a manifestly Lorentz-invariant formulation of chiral
effective field theory. By solving the coupled-channel integral equations with the full off-shell
dependence of the effective potential and applying subtractive renormalization,  we analyze the
renormalized scattering amplitudes and obtain the two-pole structure of the $\Lambda(1405)$ resonance. 
We also point out the necessity of including higher-order terms.
\end{abstract}


\maketitle

\date{\today}

\section{Introduction}

With the breakthrough in the observation of gravitational waves by the Advanced LIGO and Virgo
Collaborations~\cite{TheLIGOScientific:2017qsa}, the study of neutron stars enters a new era of
multi-message observations.  As the density increases, the strange degrees of freedom are expected to
become active in the interior of dense objects like e.g.~neutron stars. In addition to the appearance
of hyperons,  antikaon condensation can soften the equation of state and modify the bulk properties of neutron
stars~\cite{Kaplan:1986yq,Pal:2000pb}. Besides,  bound systems of antikaonic and multi-antikaonic
nuclei have been studied e.g. in Ref.~\cite{Gazda:2007wd}.  
Those phenomena attract attention to
studying the dynamics of the strong  $\bar{K}N$ interaction.
Such investigations are also essential to deepen our understanding of the SU(3) dynamics in
nonperturbative QCD.

The interaction of $\bar{K}N$ is rather strong
as manifested in the existence of the
$\Lambda(1405)$ resonance~\cite{Zyla:2020zbs} close to the $\bar{K}N$ threshold. The
$\Lambda(1405)$ resonance was first predicted by Daliz and Tuan~\cite{Dalitz:1959dn},
and it has been confirmed soon after by  hydrogen bubble chamber experiments in the analysis of
the $\pi\Sigma$ mass spectrum~\cite{Alston:1961zzd,Bastien:1961zz}. 
Since its discovery many studies have been carried out
to
uncover the nature of the $\Lambda(1405)$ resonance.
From the experimental side,
a large amount of $K^-p$ data are available such as the total cross sections
for the processes $K^-p\to\{K^-p,\,\bar{K}^0n,\,\pi^0\Sigma^0,\,\pi^+\Sigma^-,\,\pi^-\Sigma^+\}$~\cite{Humphrey:1962zz,Watson:1963zz,Sakitt:1965kh,Ciborowski:1982et},
the ratios of $K^-p$ capture rates~\cite{Tovee:1971ga,Nowak:1978au}, and the measurement of the
characteristics of kaonic hydrogen constraining the $K^-p$ $S$-wave scattering length in the
SIDDHARTA experiment~\cite{Bazzi:2011zj}. The $\Lambda(1405)$ has been
investigated within a variety of theoretical approaches, e.g.,
the relativistic quark model~\cite{Darewych:1985dc}, QCD sum rules~\cite{Kisslinger:2009dr},
phenomenological potential models~\cite{Siegel:1988rq,Fink:1989uk,Cieply:2011nq,Cieply:2015pwa,Miyahara:2018onh},
the Skyrme model~\cite{Ezoe:2020piq}, the Hamiltonian effective field theory~\cite{Liu:2016wxq}, and
the chiral unitary approach~\cite{Kaiser:1995eg,Oset:1997it, Oller:2000fj,Lutz:2001yb, Hyodo:2002pk, GarciaRecio:2002td, Jido:2003cb, Borasoy:2005ie, Borasoy:2006sr, Oller:2006jw, Ikeda:2011pi, Ikeda:2012au, Mai:2012dt, Guo:2012vv, Cieply:2016jby, Kamiya:2016jqc, Sadasivan:2018jig, Feijoo:2018den}. Detailed discussions can be found
in recent review articles, see, e.g., Refs.~\cite{Hyodo:2011ur,Gal:2016boi,Tolos:2020aln,Mai:2020ltx},
and in the review section of PDG~\cite{Zyla:2020zbs}. Besides, the $\Lambda(1405)$ resonance has also been
studied in lattice QCD simulations~\cite{Menadue:2011pd,Engel:2013ig,Hall:2014uca}. 

Among the above mentioned theoretical methods, the chiral unitary
approach that relies on chiral
perturbation theory~\cite{Weinberg:1978kz,Gasser:1984gg}
retains a special place 
as it
incorporates
important constraints from the chiral symmetry on the dynamical generation
of $\Lambda(1405)$. The most-interesting phenomenon of the two-pole structure of $\Lambda(1405)$, i.e.
that two poles are found on the same Riemann sheet, was first reported in Ref.~\cite{Oller:2000fj}.
The origin of this two-pole structure is attributed to the two attractive channels ($\pi\Sigma$ and
$\bar{K}N$) in the SU(3) basis.  Details of the two-pole structure can be found in the dedicated
review article~\cite{Meissner:2020khl}. Various studies have revealed
that the higher pole (i.e.~the one with the larger 
real part) is slightly below the threshold of $\bar{K}N$ with the
narrow width of the order of $10$ MeV. However, for the lower pole of $\Lambda(1405)$, there is
about $50$ MeV uncertainty in the real and imaginary parts obtained in different
works~\cite{Ikeda:2011pi,Ikeda:2012au,Mai:2012dt,Guo:2012vv,Cieply:2016jby}, 
because the current experimental data are not very sensitive to the lower pole. This  raised
a debate on the
one-pole versus two-pole structure of $\Lambda(1405)$ state, see,
e.g.,
Refs.~\cite{Dong:2016auh,Revai:2017isg,Myint:2018ypc,Bruns:2019bwg,Anisovich:2020lec}.
Note
that the mentioned studies finding only
a single pole have some deficiencies as detailed e.g.~in Ref.~\cite{Meissner:2020khl}.
Notice further that in chiral unitary models, the scattering amplitudes depend on the momentum cutoff
parameter $(\Lambda)$ \cite{Oset:1997it} or subtraction
constant(s)~\cite{Oller:2000fj,Borasoy:2006sr,Hyodo:2011ur}
introduced to deal with the
ultraviolet divergences in the unitarization procedure.  This results
in some model dependence of
the pole position(s) of the $\Lambda(1405)$.  

Recently we have proposed a renormalizable approach to study meson-baryon scattering by
utilizing  time-ordered perturbation theory for a manifestly Lorentz-invariant formulation of chiral
perturbation theory~\cite{Ren:2020wid}. Effective potentials are
defined as the sums of all possible two-particle
irreducible time-ordered diagrams, and the integral equations for the meson-baryon scattering
amplitudes are derived in time-ordered perturbation
theory. Renormalized amplitudes are obtained by applying
the subtractive renormalization to the solutions of the integral
equations at  leading order (LO), while  higher-order corrections are
included perturbatively in a similar fashion to 
Refs.~\cite{Epelbaum:2020maf,Ren:2019qow,Baru:2019ndr}. 
As shown in Ref.~\cite{Ren:2020wid}, our approach can be successfully applied to the pion-nucleon system.
In the current work, we apply this framework to study meson-baryon systems with strangeness $S=-1$
at LO and investigate the nature of the S-wave $\Lambda(1405)$
resonance.  This study should be considered as a first
step. In the future, we plan to extend it to include  higher-order
corrections and further experimental data 
in order to sharpen our conclusions.

The manuscript is organized as follows. In Sect.~\ref{sec:theo} we lay out the formalism to study
meson-baryon scattering in SU(3) unitarized chiral effective field theory based on a renormalizable
approach. Our results are presented and discussed in Sect.~\ref{sec:res}. We end with a summary and
outlook in Sect.~\ref{sec:sum}.

\section{Theoretical Framework}
\label{sec:theo}
In this section we briefly outline the theoretical framework of our work, which is the three-flavor
extension of the SU(2) formalism developed in  Ref.~\cite{Ren:2020wid}.

\subsection{Meson-baryon scattering amplitude}
The on-shell amplitude of the elastic meson-baryon scattering process $M_1(q_1)+B_1(p_1)\to M_2(q_2)+B_2(p_2)$
can be parameterized as 
\begin{eqnarray}
T_{M B} &=& \bar{u}_{B_2}(p_2,s_2) \left[A+\frac{1}{2}(\slashed q_1 + \slashed q_2) B\right] u_{B_1}(p_1,s_1)
\nonumber\\
&=& \bar{u}_{B_2}(p_2,s_2) \left[D+\frac{i}{m_{B_1}+m_{B_2}} \, \sigma^{\mu\nu}q_{2,\mu} q_{1,\nu} B\right]
u_{B_1}(p_1,s_1),
\end{eqnarray}
with $\sigma^{\mu\nu}=\frac{i}{2}[\gamma^\mu,\gamma^\nu]$, and $D=A+ B (s-u)/(2(m_{B_1}+m_{B_2}))$.
The conventional Mandelstam variables are defined as $s=(p_1+q_1)^2$, $t=(p_1-p_2)^2$, and
$u=(p_1-q_2)^2$ with $s+t+u=M_{M_1}^2 + m_{B_1}^2 + M_{M_2}^2 + m_{B_2}^2$. The Dirac
spinor $u(p,s)$ of a baryon is normalized according to 
\begin{equation}
u_B(p,s)=\sqrt{\frac{\omega_B(p)+m}{2 m}}\left(\begin{array}{c}
1 \\
\frac{\bm{\sigma} \cdot \bm{p}}{\omega_B(p)+m}
\end{array}\right)\chi_s,
\end{equation}
where $\chi_s$ is a two-component spinor with spin $s$, and $\omega_B(p)=\sqrt{\bm{p}^2+m^2}$ is the
energy. Following Ref.~\cite{Ren:2020wid}, we decompose the Dirac spinor as 
\begin{equation}
u_B(p)=u_{0}+\left[u(p)-u_{0}\right] \equiv \left(\begin{array}{c}
1 \\
0
\end{array}\right) \chi_s +u_{\mathrm{ho}},
\end{equation}
and consider $u_{\mathrm{ho}}$ as a higher-order contribution. 
Using the leading approximation for the Dirac spinor we obtain the reduced amplitude which reads 
\begin{eqnarray}
T_{M B} &=& D - \frac{\bm{q}_1\cdot\bm{q}_2}{m_{B_1}+m_{B_2}} B + \frac{(\bm{\sigma}\cdot\bm{q}_2)
(\bm{\sigma}\cdot\bm{q}_1)}{m_{B_1}+m_{B_2}} B	\nonumber\\[0.5em]
&\equiv& T_{MB}^{c} + i\,  \bm{\sigma}\cdot (\bm{q}_2\times\bm{q}_1)\,T_{MB}^{so},
\end{eqnarray}
where we have introduced
the non-spin-flip amplitude $T_{MB}^{c}$ and the spin-flip amplitude $T_{MB}^{so}$.

\subsection{The leading-order effective Lagrangian and  the meson-baryon interaction potential}

Below we present the LO potential of the meson-baryon interaction in the strangeness $S=-1$ sector
and the corresponding scattering equation obtained using time-ordered perturbation theory within
a manifestly Lorentz-invariant chiral effective Lagrangian.   The details of the formalism can be
found in Ref.~\cite{Ren:2020wid}. 
To keep the paper self-contained we provide below the basic ingredients of the three-flavor
extension used here.

One essential feature of our framework is that it incorporates the fields corresponding to
lowest-lying vector mesons as dynamical degrees of freedom of the effective Lagrangian.
In this formulation, the Weinberg-Tomozawa term in the
effective meson-baryon  Lagrangian is saturated by the
vector-meson exchange, which has a better ultraviolet behavior.  
The LO chiral Lagrangian used in our calculations has the form
\begin{eqnarray}\label{Eq:Lag0}
{\cal L}_\mathrm{LO} &=& \frac{F_0^2}{4}\, \left\langle u_\mu
u^\mu +\chi_+ \right\rangle  
+ \left\langle \bar{B} \left( i\gamma_\mu \partial^\mu -m \right)  B \right\rangle + \frac{D/F}{2}
 \left\langle \bar{B} \gamma_\mu \gamma_5 [u^\mu,B]_{\pm}\right\rangle   \nonumber\\
&&- \frac{1}{4} \, \left\langle  V_{\mu\nu}   V^{\mu\nu} - 2 M_V^2 \,\left( V_{\mu} -\frac{i}{g} \, \Gamma_\mu\right)  \left( V^{\mu} -\frac{i}{g} \, \Gamma^\mu\right)  \right\rangle   + g \,
\left\langle \bar{B} \gamma_\mu [V^\mu,B]\right\rangle,
\end{eqnarray}
where $\langle \ldots \rangle$ denotes the trace  in the flavor space, and 
\begin{eqnarray}
u_{\mu}&=& i (u^{\dagger} \partial_{\mu}u - u\partial_{\mu}u^{\dagger}), ~ 
~u=\exp \left( i\, P /(\sqrt{2}F_{0}) \right), \nonumber\\
\chi_{+} &=& u^{\dagger} \chi u^{\dagger} + u \chi^{\dagger} u, ~\mathrm{with}~ \chi=2 B_{0} \mathcal{M}, \nonumber\\
\Gamma_{\mu} &=& \frac{1}{2}\left(u^{\dagger} \partial_{\mu} u+u \partial_{\mu} u^{\dagger}\right), \nonumber\\
V_{\mu \nu} &=& \partial_{\mu} V_{\nu}-\partial_{\nu} V_{\mu} - ig[V_\mu,V_\nu],
\end{eqnarray}
with the pion decay constant $F_0$ in the three-flavor chiral limit, and  $D$ and $F$ - the axial-vector
couplings. We use the coupling constant of the vector-field self-interaction $g$ determined via the
KSFR relation, $M_V^2=2g^2F_0^2$. $\mathcal{M}$ denotes the quark-mass matrix and $B_0$ is related to the
scalar quark condensate, $m$ and $M_V$ stand for the octet baryon and the vector meson masses in the chiral
limit, respectively. The SU(3) matrices collecting the pseudoscalar mesons, the octet baryons and
the vector mesons are given, respectively, by
\begin{eqnarray}
P &=& \left(\begin{array}{ccc}
\frac{1}{\sqrt{2}} \pi^{0}+\frac{1}{\sqrt{6}} \eta & \pi^{+} & K^{+} \\
\pi^{-} & -\frac{1}{\sqrt{2}} \pi^{0}+\frac{1}{\sqrt{6}} \eta & K^{0} \\
K^{-} & \bar{K}^{0} & -\frac{2}{\sqrt{6}} \eta
\end{array}\right), \nonumber\\ 
B&=&\left(\begin{array}{ccc}
\frac{1}{\sqrt{2}} \Sigma^{0}+\frac{1}{\sqrt{6}} \Lambda & \Sigma^{+} & p \\
\Sigma^{-} & -\frac{1}{\sqrt{2}} \Sigma^{0}+\frac{1}{\sqrt{6}} \Lambda & n \\
\Xi^{-} & \Xi^{0} & -\frac{2}{\sqrt{6}} \Lambda
\end{array}\right),\nonumber\\
V_{\mu}&=&\left(\begin{array}{ccc}
\frac{\rho^{0}}{\sqrt{2}}+\frac{\omega}{\sqrt{2}} & \rho^{+} & K^{*+} \\
\rho^{-} & -\frac{\rho^{0}}{\sqrt{2}}+\frac{\omega}{\sqrt{2}} & K^{* 0} \\
K^{*-} & \bar{K}^{* 0} & \phi
\end{array}\right)_\mu.
\end{eqnarray} 

\begin{figure}[t]
\includegraphics[width=0.9\textwidth]{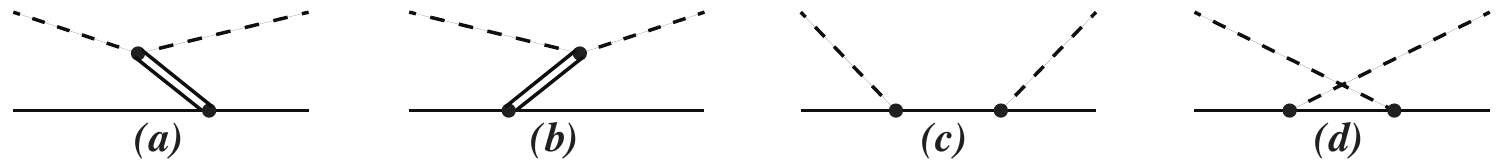}
\caption{Time-ordered diagrams contributing to
  the LO meson-baryon potential. 
  The dashed, solid and double-solid lines correspond to pseudoscalar mesons, octet baryons and
  vector mesons, respectively.}
\label{Fig:LOPot}
\end{figure}

At leading order, the meson-baryon scattering amplitude for the
process $M_i(q_1)+B_i(p_1) \to M_j(q_2)+B_j(p_2)$
is given by the time-ordered diagrams shown in
Fig.~\ref{Fig:LOPot}. Besides the Born and
crossed-Born diagrams, there are also
the vector meson exchange contributions that replace the
Weinberg-Tomozawa contact term. Notice that the contributions
stemming from  the second term in the vector meson propagator $\propto g^{\mu\nu}-q^\mu q^\nu/M_V^2$
are suppressed compared to the ones of the
first term for considered low-energy processes and need to be taken into account together with other
higher-order corrections. 
We further emphasize that possible power-counting-breaking contributions stemming from the decays $V\to PP$~\cite{Bruns:2004tj} in loop diagrams can be absorbed in vertex corrections~\cite{Kubis:2000zd}. 


\begin{table}[b]
\caption{Coefficients $C_{M_jB_j,M_iB_i}^V$ of the vector meson exchange contributions in the
    strangeness $S=-1$ sector with isospin $I=0$.}
\label{Tab:cVij}
\begin{tabular}{l|c|c|c|c}
\hline 
 $C^V$ & $\pi\Sigma$ & $\bar{K}N$ & $\eta\Lambda$ & $K\Xi$ \\
  \hline
 $\pi\Sigma$ & $C^\rho=-16$ & $C^{K^*}=2\sqrt{6}$ & 0 & $C^{K^*}=-2\sqrt{6}$ \\  
 $\bar{K}N$ &  & $C^{\{\rho,\omega,\phi\}}=\{-6,-2,-4\}$ & $C^{K^*}=-6\sqrt{2}$ & 0 \\  
 $\eta\Lambda$ & & & 0 & $C^{K^*}=6\sqrt{2}$ \\
 $K\Xi$ & & & & $C^{\{\rho,\omega,\phi\}}=\{-6,-2,-4\}$ \\
 \hline 
\end{tabular}
\end{table}

\begin{table}[t]
\caption{Coefficients $C_{M_jB_j,M_iB_i}^B$ of the Born terms in the strangeness $S=-1$ sector
    with isospin $I=0$.}
\label{Tab:cBij}
\begin{tabular}{l|c|c|c|c}
\hline 
 $C^{B}$ & $\pi\Sigma$ & $\bar{K}N$ & $\eta\Lambda$ & $K\Xi$ \\
  \hline
 $\pi\Sigma$ & $C^\Lambda=4D^2$ & $C^\Lambda=\sqrt{\frac{8}{3}}D(D+3F)$ & $C^\Lambda=\frac{4}{\sqrt{3}}D^2$ & $C^\Lambda=-\sqrt{\frac{8}{3}}D(D-3F)$ \\  
 $\bar{K}N$ &  & $C^\Lambda=\frac{2}{3}(D+3F)^2$ & $C^\Lambda=\frac{2\sqrt{2}}{3}D(D+3F)$ & $C^\Lambda=-\frac{2}{3}(D^2-9F^2)$ \\  
 $\eta\Lambda$ & & & $C^\Lambda=\frac{4}{3}D^2$ & $C^\Lambda=-\frac{2\sqrt{2}}{3}D(D-3F)$ \\
 $K\Xi$ & & & & $C^\Lambda=\frac{2}{3}(D-3F)^2$ \\
 \hline 
\end{tabular}
\end{table}

\begin{table}[t]
\caption{Coefficients $\tilde{C}_{M_jB_j,M_iB_i}^B$ of the crossed-Born terms in the strangeness $S=-1$
    sector with isospin $I=0$.}
\label{Tab:tcBij}
\vspace{3mm}
\begin{tabular}{l|c|c|c|c}
\hline 
 $\tilde{C}^{B}$ & $\pi\Sigma$ & $\bar{K}N$ & $\eta\Lambda$ & $K\Xi$ \\
  \hline
 $\pi\Sigma$ & $\tilde{C}^{\{\Lambda,\Sigma\}}=\{\frac{4D^2}{3}, -8F^2\}$ & $\tilde{C}^N=\sqrt{6}(F^2-D^2)$ & $\tilde{C}^\Sigma=-\frac{4}{\sqrt{3}}D^2$ & $\tilde{C}^\Xi=\sqrt{6}(D^2-F^2)$ \\  
 $\bar{K}N$ &  & $0$ & $\tilde{C}^N=\frac{\sqrt{2}(D^2-9F^2)}{3}$ & $\tilde{C}^{\{\Lambda,\Sigma\}}=\{\frac{9F^2-D^2}{3},3(F^2-D^2)\}$ \\  
 $\eta\Lambda$ & & & $\tilde{C}^\Lambda=\frac{4}{3}D^2$ & $\tilde{C}^\Xi=\frac{\sqrt{2}}{3}(9F^2-D^2)$ \\
 $K\Xi$ & & & & 0 \\
 \hline 
\end{tabular}
\end{table}

In this LO calculation we employ the scattering amplitudes in the isospin limit using the
averaged masses for the mesons and baryons. The LO potential in the isospin formalism is given by
\begin{eqnarray}\label{Eq:LOpot}
  V_{M_jB_j,M_iB_i}^{(a+b)} &=& -\frac{1}{32 F_0^2} \sum\limits_{V} C_{M_jB_j,M_iB_i}^V \frac{M_V^2}{\omega_V(q_1-q_2)} \left(\omega_{M_i}(q_1)+\omega_{M_j}(q_2)\right) \nonumber\\
  && \times \left(\frac{1}{E-\omega_{B_i}(p_1)-\omega_V(q_1-q_2)-\omega_{M_j}(q_2)} \right. \nonumber\\
  && \left. \quad + \frac{1}{E-\omega_{B_j}(p_2)-\omega_V(q_1-q_2)-\omega_{M_i}(q_1)}  \right),\nonumber\\
  V_{M_jB_j,M_iB_i}^{(c)} &=& \frac{1}{4F_0^2}\sum\limits_{B} C_{M_jB_j,M_iB_i}^{B} \frac{m_{B}}{\omega_{B}(P)} \frac{(\bm{\sigma}\cdot\bm{q}_2)(\bm{\sigma}\cdot\bm{q}_1)}{E-\omega_{B}(P)}  ,\nonumber\\
   V_{M_jB_j,M_iB_i}^{(d)} &=& \frac{1}{4F_0^2}\sum\limits_{B} \tilde{C}_{M_jB_j,M_iB_i}^{B} \frac{m_{B}}{\omega_{B}(K)} \frac{(\bm{\sigma}\cdot\bm{q}_1)(\bm{\sigma}\cdot\bm{q}_2)}{E-\omega_{M_i}(q_1)-\omega_{M_j}(q_2)-\omega_{B}(K)} ,
\end{eqnarray}
where the expressions are summed over
all vector mesons $V$ in the vector-meson-exchange contributions,
and over the internal baryons $B$ in the Born and crossed-Born terms. Further, $E = \sqrt{s}$ is the total
energy of the meson-baryon system, and 
$P=q_1+p_1=q_2+p_2$, $K=p_1-q_2=p_2-q_1$. The on-shell energy of a
particle is given by $\omega_{X}(p)
\equiv \sqrt{m_X^2 + \bm{p}^2}$.

In the $S=-1$ sector, there are four coupled channels with isospin
$I=0$, namely $\pi\Sigma$, $\bar{K}N$, $\eta\Lambda$ and $K\Xi$.
The various coefficients $C_{M_jB_j,M_iB_i}^{V}$, $C_{M_jB_j,M_iB_i}^{B}$ and
$\tilde{C}_{M_jB_j,M_iB_i}^{B}$ are tabulated  in Tables~\ref{Tab:cVij}, \ref{Tab:cBij} and \ref{Tab:tcBij},
where the indices $i,~j$ represent the particle channels. 
Here, we use the phase convention  $\left|\pi^{+}\right\rangle=-|1,1\rangle,~\left|K^{-}\right\rangle
=-|1 / 2,-1 / 2\rangle,~\left|\Sigma^{+}\right\rangle=$ $-|1,1\rangle$ and $\left|\Xi^{-}\right\rangle
=-|1 / 2,-1 / 2\rangle$ for the isospin states.

We rewrite the LO  potential as the central and spin-orbital parts, 
\begin{eqnarray}
  V_{M_jB_j,M_iB_i} &=& V_{M_jB_j,M_iB_i}^{(a+b)} + V_{M_jB_j,M_iB_i}^{(c)} + V_{M_jB_j,M_iB_i}^{(d)} \nonumber\\
  &\equiv &  W^c_{M_jB_j,M_iB_i} + i\, \bm{\sigma}\cdot (\bm{q}_2\times \bm{q}_1) W^{so}_{M_jB_j,M_iB_i}.  
\end{eqnarray}
It is convenient to calculate the amplitude in the center-of-mass (CMS) frame with 
\begin{equation}
\begin{aligned}
q_{1}^\mu &=\left(\omega_{M_i}(\bm{p}), \bm{p}\right),\quad p_{1}^\mu
=\left(\omega_{B_i}(\bm{p}),-\bm{p}\right),\quad  \\
q_{2}^\mu &=\left(\omega_{M_j}(\bm{p}'), \bm{p}'\right),\quad p_{2}^\mu
=\left(\omega_{B_j}(\bm{p}'),-\bm{p}'\right),
\end{aligned}
\end{equation}
where the relative momenta $\bm{p}$ and $\bm{p}'$ are introduced. A state of the meson-baryon system
can be represented as a sum of partial wave states, which are denoted as $|LJ\rangle$, 
where  $J$ is the total angular momentum and $L=J\pm 1/2$ is the orbital angular momentum. Because of
parity conservation in the strong interactions, the $L_+=J-1/2$ and the $L_- = J+1/2$ waves are obviously
decoupled.   
The partial wave projection of the potential in the isospin basis is given by 
\begin{eqnarray}
   &&\langle L_{\pm} J| V_{M_jB_j,M_iB_i} |L_{\pm} J\rangle
    \equiv V_{M_jB_j,M_iB_i}^{LJ} = \nonumber\\
  &&  2\pi \int_{-1}^1 dz \, \left[ W^c_{M_jB_j,M_iB_i} P_{L_\pm}(z) + p^2 W^{so}_{M_jB_j,M_iB_i} P_{L_{\pm} \pm 1 }(z) - z p^2 W^{so}_{M_jB_j,M_iB_i} P_{L_{\pm}}(z) \right],
\end{eqnarray}
where $z=\cos\theta$, with $\theta$ the angle between $\bm{p}$ and
$\bm{p}'$, $p \equiv | \bm{p} |$ and $P_L(z)$ denotes the
Legendre polynomial.

\subsection{Partial wave integral equations and subtractive renormalization} 
\begin{figure}[t]
\includegraphics[width=0.9\textwidth]{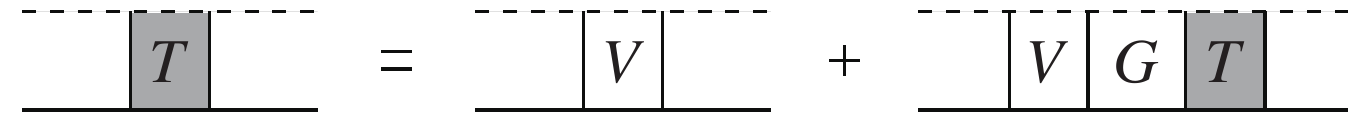}
\caption{Diagrammatic representation of the meson-baryon scattering equation. The dashed and
  solid lines denote the pseudoscalar mesons and the octet baryons, respectively.}
\label{Fig:IntEq}
\end{figure}
In time-ordered perturbation theory one obtains the coupled-channel integral equation for the
$T$-matrix, which is visualized in Fig.~\ref{Fig:IntEq},  
\begin{eqnarray}\label{Eq:inteq}
T_{M_jB_j,M_iB_i}(\bm{p}',\bm{p}; E) &=& V_{M_jB_j,M_iB_i}(\bm{p}',\bm{p}; E)   \nonumber\\
& +& \sum_{MB}
\int\frac{d ^3\bm{k}}{(2 \pi)^3} V_{M_jB_j,MB}(\bm{p}',\bm{k}; E)\, G_{MB}(E) \, T_{MB,M_iB_i}(\bm{k},\bm{p}; E),
\end{eqnarray}
where $M_iB_i,M_jB_j$ and $MB$ denote the initial, final and intermediate
particle channels, and the two-body Green functions read
\begin{equation}\label{Eq:Gij}
G_{MB}(E)= \frac{1}{2\, \omega_M \omega_B}\, \frac{m_B}{E - \omega_M-\omega_B+i \epsilon} \,.
\end{equation}
Projecting onto specific partial waves in the $|LJ\rangle$ basis, the integral equation is written as
\begin{eqnarray}\label{Eq:pwinteq}
T_{M_jB_j,M_iB_i}^{LJ}(p',p; E) &=& V_{M_jB_j,M_iB_i}^{LJ}(p',p; E)   \nonumber\\
& +& \sum_{MB}
\int\frac{dk k^2}{(2 \pi)^3} \, V_{M_jB_j,MB}^{LJ}(p',k; E)\, G_{MB}(E) \, T_{MB,M_iB_i}^{LJ}(k,p; E),
\end{eqnarray}
where $p,~p',~k$ are defined as the magnitudes of the momenta,  $p=|\bm{p}|$, $p'=|\bm{p}'|$, $k=|\bm{k}|$.
Since the LO potential can be divided into the one-baryon-reducible and irreducible parts, 
\begin{equation}
  V_{M_jB_j,M_iB_i} = V^R_{M_jB_j,M_iB_i} + V^I_{M_jB_j,M_iB_i},	
\end{equation}
with $V^R=V^{(c)}$ and $V^I=V^{(a+b)}+V^{(d)}$,  we can apply a subtractive renormalization to
obtain the finite on-shell $T$-matrix. 
Using a symbolic notation, the meson-baryon integral equation $T=V+VGT$ can be rewritten as a system of coupled
equations as
\begin{eqnarray}
  T&=& T_I + (1+T_I\,G)\,T_R\,(1+G\,T_I),\nonumber\\
  T_I &=& V_I + V_I\, G\, T_I,\nonumber\\
  T_R &=& V_R + V_R\, G\,(1+T_I\,G)\,T_R. 	
\end{eqnarray}
In order to obtain the renormalized finite T-matrix,  
we replace the meson-baryon propagator $G_{MB}(E)$ with the subtracted one $G_{MB}^S(E)=G_{MB}(E)-G_{MB}(m_B)$.
This corresponds to including the contributions of an infinite number of meson-baryon counter-terms 
(details can be found in Refs.~\cite{Epelbaum:2020maf,Ren:2020wid}).  As discussed in
Ref.~\cite{Ren:2020wid}, the subtractive renormalization does not affect the dynamics of bound
states or resonances. 


\section{Results and discussion}
\label{sec:res}


In the numerical evaluation, the masses of pesudoscalar mesons are taken as $M_\pi=138.0$~MeV,
$M_K=495.6$~MeV, and $M_\eta=547.9$~MeV. For the masses of octet baryons we use
$m_N=938.9$~MeV, $m_\Lambda=1115.6$~MeV, $m_\Sigma=1193.1$~MeV and
$m_\Xi=1318.0$~MeV, while for 
the vector mesons the values $M_\rho=775.3$~MeV, $M_{K^*}=893.7$~MeV, $M_\omega=782.7$~MeV, and
$M_\phi=1019.5$~MeV~\cite{Zyla:2020zbs} are employed. The axial vector couplings are taken as $D=0.760$ and $F=0.507$,
with $D+F=g_A=1.267$. The pseudoscalar decay constants are fixed as $F_\pi=92.07$ MeV, $F_K=110.1$ MeV
and $F_\eta\approx 1.2F_\pi$  according to the PDG average values~\cite{Zyla:2020zbs}.  
Note that at this order, there is no free parameter (see also the discussion below). Thus,
at this order we have pure predictions.

%

\begin{table}[t]
\centering
\caption{Pole positions $z_R$ of $\Lambda(1405)$ in the $S=-1$ sector (units are MeV).}
\label{Tab:1405}
\begin{tabular}{p{3cm}p{3cm}p{4cm}p{4cm}}
\hline\hline
 &  & lower pole & higher pole \\
  \hline
 \multirow{2}{*}{\begin{minipage}{0.6in}This~work (LO)\end{minipage} } &$F_0=F_\pi$  & $1337.7 - i\, 79.1$ & $1430.9 - i\, 8.0$ \\
   & $F_0=103.4$   & $1348.2-i\,120.2$ & $1436.3-i\,0.7$ \\
 \hline 
 \multirow{4}{*}{NLO} 
 & Ref.~\cite{Ikeda:2011pi,Ikeda:2012au}  & $1381_{-6}^{+18}-i 81_{-8}^{+19}$ & $1424_{-23}^{+7}-i 26_{-14}^{+3}$ \\ 
& Ref.~\cite{Guo:2012vv}, Fit II  & $1388_{-9}^{+9}-i 114_{-25}^{+24}$ & $1421_{-2}^{+3}-i 19_{-5}^{+8}$ \\ 
 
& Ref.~\cite{Mai:2014xna}, sol-2  & $1330_{-5}^{+4}-i 56_{-11}^{+17}$ & $1434_{-2}^{+2}-i 10_{-1}^{+2}$ \\
& Ref.~\cite{Mai:2014xna}, sol-4  & $1325_{-15}^{+15}-i 90_{-18}^{+12}$ & $1429_{-7}^{+8}-i 12_{-3}^{+2}$ \\
 \hline\hline 
\end{tabular}
\end{table}
 
To obtain the $I=0$ $\bar{K}N$ scattering $T$-matrix with coupled-channel effects taken into account
we need to solve the integral equation, Eq.~(\ref{Eq:pwinteq}), with the four coupled channels
$\bar{K}N$, $\pi\Sigma$, $\eta\Lambda$, and $K\Xi$. Instead of introducing the approximation of the
on-shell factorization (see, e.g.,~Ref.~\cite{Hyodo:2011ur}), we solve the scattering equation with
the full off-shell dependence. As argued in Refs.~\cite{Bruns:2010sv, Mai:2012dt,Morimatsu:2019wvk}, this treatment is
necessary since the pole positions of resonances can change in solving the Bethe-Salpeter equation with
the full off-shell dependence of the chiral potential. Then, using the S-wave projection of the chiral
LO potential of Eq.~(\ref{Eq:LOpot}), we employ the subtractive renormalization to obtain the finite
S-wave $T$-matrix that does not contain contributions increasing with
the cutoff $\Lambda$, i.e.~one can take the limit
$\Lambda\to \infty$.\footnote{In practice, we prefer to choose 
  finite cutoff values like $\Lambda=10$~GeV that are sufficiently
  large to keep finite-$\Lambda$ artifacts negligibly small.}
It is worth noticing that our framework does not have the usual obstacle of the large cutoff-dependence
of the $T$-matrix present in the traditional chiral unitary approach.   

Performing an analytic continuation of the $T$-matrix into the complex $s$-plane, we find the $\Lambda(1405)$
resonance with the two-pole structure in the second Riemann sheet with only the $\pi\Sigma$ channel
open for decay (i.e.~$(M_\pi+m_\Sigma)^2 < s < (M_{\bar{K}}+m_N)^2$).  The obtained pole positions
are denoted as the ``lower'' pole and ``higher'' pole  and listed in Table~\ref{Tab:1405}. By varying the
meson-decay constant, $F_0$, from the physical SU(2) value $F_\pi=92.07$ MeV to its SU(3)-average
value $103.4$~MeV, we find that the width of the first pole is increasing. The second pole lies close
and moves beyond the threshold of $\bar{K}N$ channel, and its width decreases, when the meson decay
constant increases. Our LO results are consistent with the ones of the NLO study of Ref.~\cite{Mai:2014xna},
in particular, for what concerns the results for the lower pole.

\begin{table}[t]
\caption{The (absolute) values of the coupling strengths $g_i$ ($|g_i|$) for the two poles
    of $\Lambda(1405)$ resonance.}
\label{Tab:gij}
\vspace{3mm}
\begin{tabular}{p{2cm}p{3cm}p{2cm}p{0.5cm}p{3cm}p{2cm}}
\hline\hline
     & \multicolumn{2}{c}{lower pole} & & \multicolumn{2}{c}{higher pole}  \\
\cline{2-3} \cline{5-6}
  & $g_i$ & $|g_i|$ & & $g_i$ & $|g_i|$ \\
\hline 
$\pi\Sigma$ & $1.83+i1.90$ & $2.64$ &  &  $-0.38+i0.84$ & $0.92$   \\
$\bar{K}N$ & $-1.59-i1.47$ & $2.17$ & & $2.16-i0.83$ & $2.31$ \\ 
$\eta\Lambda$ & $-0.19-i0.67$ & $0.69$ & & $1.59-i0.36$ & $1.63$
\\
$K\Xi$ & $0.72+i0.81$ & $1.08$ & & $-0.10+i0.34$ & $0.35$ \\
\hline\hline
\end{tabular}	
\end{table} 
It is further interesting to investigate the structure of the above two poles.  Approaching the pole
position $z_R$, the on-shell scattering $T$-matrix can be approximated by
\begin{equation}
	T_{ij} \simeq 4\pi  \frac{g_{i} \, g_{j}}{z-z_{R}},
\end{equation}
where $g_{i}\, (g_j)$ represents the contribution to the coupling strength of the initial (final)
transition channel. In general, the couplings $g_{i},\, g_j$, which can be extracted from  the
residues of the $T$-matrix, are complex-valued numbers.
The couplings obtained for the $\Lambda(1405)$ resonance are tabulated in Table~\ref{Tab:gij}. One
can see that the lower pole couples predominantly to the $\pi\Sigma$ channel, while the higher
pole couples strongly to the $\bar{K}N$ channel. This could explain the large imaginary part of the
lower pole, which is the consequence of the strong $\pi\Sigma$ coupling. Thus, the different coupling
nature to the meson-baryon channels form the two-pole structure of $\Lambda(1405)$.  

\begin{figure}[t] 
\includegraphics[width=0.6\textwidth]{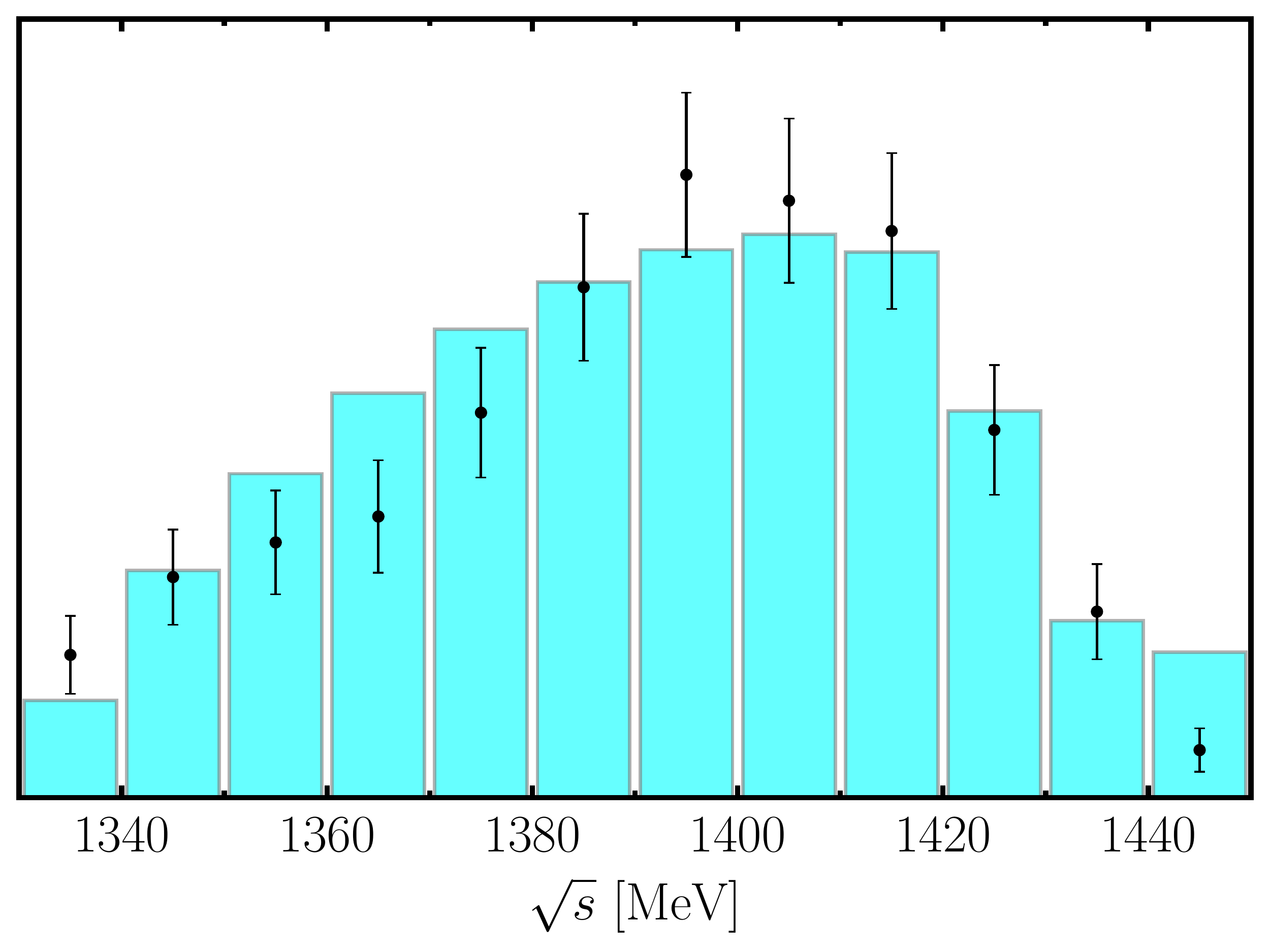}
\caption{$\pi\Sigma$ invariant mass spectrum with $I=0$ in arbitrary
  units.
  The histogram represents the result of our calculations as explained
  in the text. Experimental data for the $\pi^-\Sigma^+$ event
  distribution are taken from 
  Ref.~\cite{Hemingway:1984pz}.} 
\label{Fig:invmasspiSigma}
\end{figure}

Furthermore, we present the shape of the $\Lambda(1405)$ spectrum in Fig.~\ref{Fig:invmasspiSigma},
where the calculated invariant mass distribution is compared with 
the experimental data of $\pi^-\Sigma^+$ channel
from~Ref.~\cite{Hemingway:1984pz}. The event distribution is
calculated by taking into account both the  $\pi\Sigma\to\pi\Sigma$
and $\bar{K}N\to\pi\Sigma$ channels as described in
Ref.~\cite{Oller:2000fj}.

Finally, we note that we have also calculated the $\bar{K}N$ scattering lengths for $I=0$ and $I=1$.
The isospin-zero scattering length turns out to be $a_0 = -2.50 + i
1.37$~fm, which is somewhat outside of the
region allowed by combining the scattering data with the SIDDHARTA
kaonic hydrogen result, see
e.g.~Ref.~\cite{Doring:2011xc}. We expect this issue to be resolved
upon including NLO corrections.
Our LO result for the isovector scattering length $a_1 = 0.33 + i 0.72$~fm is, on the
other hand, within the allowed region
mapped out from scattering and kaonic hydrogen data.

\section{Summary and outlook}
\label{sec:sum}
In this paper we have studied meson-baryon scattering in the strangeness $S=-1$ sector to
investigate the structure of the $\Lambda(1405)$ resonance using
time-ordered perturbation theory applied to the
Lorentz-invariant effective chiral Lagrangian with the explicit
inclusion of low-lying vector mesons. In the considered framework, the effective potential of the meson-baryon
scattering is defined as a sum of two-particle irreducible time-ordered diagrams. 
The renormalized S-wave amplitudes with $I=0$ are obtained by taking into account the full
off-shell dependence of the potential in the coupled-channel integral equations and applying
subtractive renormalization. 

In our leading-order study with no free parameters, we obtain the two-pole structure of the
$\Lambda(1405)$ state. The higher pole at $E_R = 1431-i\, 8$ MeV is mainly coupled to the $\bar{K}N$ channel,
while the lower one located at $E_R = 1338-i\, 79$ MeV couples mainly to the $\pi\Sigma$ channel.
It is worth noticing that our results are independent on the momentum
cutoff. The obtained $\pi \Sigma$ invariant mass distribution agrees
well with the experimental data while the $\bar{K}N$  isoscalar scattering
length is found to have a somewhat too large (in magnitude) real part.

The existing data and the upcoming experiments focused on
investigating the $\bar{K}N$ dynamics,
such as the lowest-energy beam of strange hadron production in JLab  experiments~\cite{Amaryan:2020xhw},
the kaonic hydrogen SIDDHARTA experiment~\cite{Bazzi:2011zj}, the photoproduction data from
JLab~\cite{Moriya:2013eb}
and the experiments with kaonic nuclear bound states
~\cite{Miliucci:2021drx,Sakuma:2020btn}
provide a strong motivation to extend our renormalizable framework to
next-to-leading order. Work along these lines is in progress.  

\section*{Acknowledgements}
This work was supported in part by BMBF (Grant No. 05P18PCFP1), by DFG and NSFC through funds provided to the
Sino-German CRC 110 ``Symmetries and the Emergence of Structure in QCD" (NSFC
Grant No.~11621131001, Project-ID 196253076 - TRR 110), by Collaborative Research Center ``The Low-Energy
Frontier of the Standard Model'' (DFG, Project No. 204404729 - SFB 1044), by the Cluster of
Excellence ``Precision Physics, Fundamental Interactions, and Structure of Matter'' (PRISMA$^+$, EXC 2118/1)
within the German Excellence Strategy (Project ID 39083149),  
by the  Georgian Shota Rustaveli National Science Foundation (Grant No. FR17-354), by VolkswagenStiftung
(Grant No. 93562), by the CAS President's International Fellowship Initiative (PIFI) (Grant No.~2018DM0034)
and by the EU (STRONG2020).


\begin{references}
	
\bibitem{TheLIGOScientific:2017qsa}
B.~P.~Abbott \textit{et al.} [LIGO Scientific and Virgo],
Phys. Rev. Lett. \textbf{119},  161101 (2017); Phys. Rev. X \textbf{9}, 011001 (2019).


\bibitem{Kaplan:1986yq}
D.~B.~Kaplan and A.~E.~Nelson,
Phys. Lett. B \textbf{175}, 57-63 (1986)

\bibitem{Pal:2000pb}
S.~Pal, D.~Bandyopadhyay and W.~Greiner,
Nucl. Phys. A \textbf{674}, 553-577 (2000).

\bibitem{Gazda:2007wd}
D.~Gazda, E.~Friedman, A.~Gal and J.~Mares,
Phys. Rev. C \textbf{76}, 055204 (2007)
[erratum: Phys. Rev. C \textbf{77}, 019904 (2008)].

\bibitem{Zyla:2020zbs}
P.~A.~Zyla \textit{et al.} [Particle Data Group],
PTEP \textbf{2020}, 083C01 (2020).

\bibitem{Dalitz:1959dn}
R.~H.~Dalitz and S.~F.~Tuan,
Phys. Rev. Lett. \textbf{2}, 425-428 (1959); Annals Phys. \textbf{10}, 307-351 (1960). 


\bibitem{Alston:1961zzd}
M.~H.~Alston, L.~W.~Alvarez, P.~Eberhard, M.~L.~Good, W.~Graziano, H.~K.~Ticho and S.~G.~Wojcicki,
Phys. Rev. Lett. \textbf{6}, 698-702 (1961).

\bibitem{Bastien:1961zz}
P.~L.~Bastien, M.~Ferro-Luzzi and A.~H.~Rosenfeld,
Phys. Rev. Lett. \textbf{6}, 702 (1961).

\bibitem{Humphrey:1962zz}
W.~E.~Humphrey and R.~R.~Ross,
Phys. Rev. \textbf{127}, 1305-1323 (1962).

\bibitem{Watson:1963zz}
M.~B.~Watson, M.~Ferro-Luzzi and R.~D.~Tripp,
Phys. Rev. \textbf{131}, 2248-2281 (1963).

\bibitem{Sakitt:1965kh}
M.~Sakitt, T.~B.~Day, R.~G.~Glasser, N.~Seeman, J.~H.~Friedman, W.~E.~Humphrey and R.~R.~Ross,
Phys. Rev. \textbf{139}, B719 (1965).

\bibitem{Ciborowski:1982et}
J.~Ciborowski, J.~Gwizdz, D.~Kielczewska, R.~J.~Nowak, E.~Rondio, J.~A.~Zakrzewski, M.~Goossens, G.~Wilquet, N.~H.~Bedford and D.~Evans, \textit{et al.}
J. Phys. G \textbf{8}, 13-32 (1982).

\bibitem{Tovee:1971ga}
D.~N.~Tovee, D.~H.~Davis, J.~Simonovic, G.~Bohm, J.~Klabuhn, F.~Wysotzki, M.~Csejthey-Barth, J.~H.~Wickens, T.~Cantwell and C.~Ni Ghogain, \textit{et al.}
Nucl. Phys. B \textbf{33}, 493-504 (1971).

\bibitem{Nowak:1978au}
R.~J.~Nowak, J.~Armstrong, D.~H.~Davis, D.~J.~Miller, D.~N.~Tovee, D.~Bertrand, M.~Goossens, G.~Vanhomwegen, G.~Wilquet and M.~Abdullah, \textit{et al.}
Nucl. Phys. B \textbf{139}, 61-71 (1978).

\bibitem{Bazzi:2011zj}
M.~Bazzi \textit{et al.} [SIDDHARTA collaboration],
Phys. Lett. B \textbf{704}, 113-117 (2011).

\bibitem{Darewych:1985dc}
J.~W.~Darewych, R.~Koniuk and N.~Isgur,
Phys. Rev. D \textbf{32}, 1765 (1985).

\bibitem{Kisslinger:2009dr}
L.~S.~Kisslinger and E.~M.~Henley,
Eur. Phys. J. A \textbf{47}, 8 (2011).

\bibitem{Siegel:1988rq}
P.~B.~Siegel and W.~Weise,
Phys. Rev. C \textbf{38}, 2221-2229 (1988).

\bibitem{Fink:1989uk}
P.~J.~Fink, Jr., G.~He, R.~H.~Landau and J.~W.~Schnick,
Phys. Rev. C \textbf{41}, 2720-2725 (1990).

\bibitem{Cieply:2011nq}
A.~Cieply and J.~Smejkal,
Nucl. Phys. A \textbf{881}, 115-126 (2012).

\bibitem{Cieply:2015pwa}
A.~Ciepl\'y and V.~Krej\v{c}i\v{r}\'\i{}k,
Nucl. Phys. A \textbf{940}, 311-330 (2015).

\bibitem{Miyahara:2018onh}
K.~Miyahara, T.~Hyodo and W.~Weise,
Phys. Rev. C \textbf{98}, 025201 (2018).

\bibitem{Ezoe:2020piq}
T.~Ezoe and A.~Hosaka,
Phys. Rev. D \textbf{102}, 014046 (2020).

\bibitem{Liu:2016wxq}
Z.-W.~Liu, J.~M.~M.~Hall, D.~B.~Leinweber, A.~W.~Thomas and J.~J.~Wu,
Phys. Rev. D \textbf{95}, 014506 (2017).

\bibitem{Kaiser:1995eg}
N.~Kaiser, P.~B.~Siegel and W.~Weise,
Nucl. Phys. A \textbf{594}, 325-345 (1995).

\bibitem{Oset:1997it}
E.~Oset and A.~Ramos,
Nucl. Phys. A \textbf{635}, 99-120 (1998).

\bibitem{Oller:2000fj}
J.~A.~Oller and U.-G.~Mei\ss ner,
Phys. Lett. B \textbf{500}, 263-272 (2001).

\bibitem{Lutz:2001yb}
M.~F.~M.~Lutz and E.~E.~Kolomeitsev,
Nucl. Phys. A \textbf{700}, 193-308 (2002).

\bibitem{Hyodo:2002pk}
T.~Hyodo, S.~I.~Nam, D.~Jido and A.~Hosaka,
Phys. Rev. C \textbf{68}, 018201 (2003).

\bibitem{GarciaRecio:2002td}
C.~Garcia-Recio, J.~Nieves, E.~Ruiz Arriola and M.~J.~Vicente Vacas,
Phys. Rev. D \textbf{67}, 076009 (2003).

\bibitem{Jido:2003cb}
D.~Jido, J.~A.~Oller, E.~Oset, A.~Ramos and U.-G.~Mei\ss ner,
Nucl. Phys. A \textbf{725}, 181-200 (2003).

\bibitem{Borasoy:2005ie}
B.~Borasoy, R.~Nissler and W.~Weise,
Eur. Phys. J. A \textbf{25}, 79-96 (2005).

\bibitem{Borasoy:2006sr}
B.~Borasoy, U.-G.~Mei\ss ner and R.~Nissler,
Phys. Rev. C \textbf{74}, 055201 (2006).

\bibitem{Oller:2006jw}
J.~A.~Oller,
Eur. Phys. J. A \textbf{28}, 63-82 (2006).

\bibitem{Ikeda:2011pi}
Y.~Ikeda, T.~Hyodo and W.~Weise,
Phys. Lett. B \textbf{706}, 63-67 (2011).

\bibitem{Ikeda:2012au}
Y.~Ikeda, T.~Hyodo and W.~Weise,
Nucl. Phys. A \textbf{881}, 98-114 (2012).

\bibitem{Mai:2012dt}
M.~Mai and U.-G.~Mei\ss ner,
Nucl. Phys. A \textbf{900}, 51 - 64 (2013).

\bibitem{Guo:2012vv}
Z.~H.~Guo and J.~A.~Oller,
Phys. Rev. C \textbf{87}, 035202 (2013).

\bibitem{Cieply:2016jby}
A.~Ciepl\'y, M.~Mai, U.-G.~Mei\ss{}ner and J.~Smejkal,
Nucl. Phys. A \textbf{954}, 17-40 (2016).

\bibitem{Kamiya:2016jqc}
Y.~Kamiya, K.~Miyahara, S.~Ohnishi, Y.~Ikeda, T.~Hyodo, E.~Oset and W.~Weise,
Nucl. Phys. A \textbf{954}, 41-57 (2016).

\bibitem{Sadasivan:2018jig}
D.~Sadasivan, M.~Mai and M.~D\"oring,
Phys. Lett. B \textbf{789}, 329-335 (2019).

\bibitem{Feijoo:2018den}
A.~Feijoo, V.~Magas and A.~Ramos,
Phys. Rev. C \textbf{99}, 035211 (2019).

\bibitem{Hyodo:2011ur}
T.~Hyodo and D.~Jido,
Prog. Part. Nucl. Phys. \textbf{67}, 55-98 (2012).

\bibitem{Gal:2016boi}
A.~Gal, E.~V.~Hungerford and D.~J.~Millener,
Rev. Mod. Phys. \textbf{88}, 035004 (2016).


\bibitem{Tolos:2020aln}
L.~Tolos and L.~Fabbietti,
Prog. Part. Nucl. Phys. \textbf{112}, 103770 (2020).

\bibitem{Mai:2020ltx}
M.~Mai,
[arXiv:2010.00056 [nucl-th]].

\bibitem{Menadue:2011pd}
B.~J.~Menadue, W.~Kamleh, D.~B.~Leinweber and M.~S.~Mahbub,
Phys. Rev. Lett. \textbf{108}, 112001 (2012).

\bibitem{Engel:2013ig}
G.~P.~Engel \textit{et al.} [BGR],
Phys. Rev. D \textbf{87}, 074504 (2013).

\bibitem{Hall:2014uca}
J.~M.~M.~Hall, W.~Kamleh, D.~B.~Leinweber, B.~J.~Menadue, B.~J.~Owen, A.~W.~Thomas and R.~D.~Young,
Phys. Rev. Lett. \textbf{114},  132002 (2015).

\bibitem{Weinberg:1978kz}
S.~Weinberg,
Physica A \textbf{96}, , 327-340 (1979).

\bibitem{Gasser:1984gg}
J.~Gasser and H.~Leutwyler,
Nucl. Phys. B \textbf{250}, 465-516 (1985).

\bibitem{Meissner:2020khl}
U.-G.~Mei{\ss}ner,
Symmetry \textbf{12}, 981 (2020).

\bibitem{Dong:2016auh} 
  F.~Y.~Dong, B.~X.~Sun and J.~L.~Pang,
  Chin.\ Phys.\ C {\bf 41}, no. 7, 074108 (2017).


\bibitem{Revai:2017isg}
J.~R\'evai,
Few Body Syst. \textbf{59}, 49 (2018).

\bibitem{Myint:2018ypc}
K.~S.~Myint, Y.~Akaishi, M.~Hassanvand and T.~Yamazaki,
PTEP \textbf{2018}, no.7, 073D01 (2018).

\bibitem{Bruns:2019bwg}
P.~C.~Bruns and A.~Ciepl\'y,
Nucl. Phys. A \textbf{996}, 121702 (2020).

\bibitem{Anisovich:2020lec}
A.~V.~Anisovich, A.~V.~Sarantsev, V.~A.~Nikonov, V.~Burkert, R.~A.~Schumacher, U.~Thoma and E.~Klempt,
Eur. Phys. J. A \textbf{56}, 139 (2020).
\bibitem{Ren:2020wid}
X.-L.~Ren, E.~Epelbaum, J.~Gegelia and U.-G.~Mei\ss{}ner,
Eur. Phys. J. C \textbf{80}, 406 (2020).

\bibitem{Epelbaum:2020maf}
E.~Epelbaum, A.~M.~Gasparyan, J.~Gegelia, U.-G.~Mei\ss{}ner and X.-L.~Ren,
Eur. Phys. J. A \textbf{56}, 152 (2020).

\bibitem{Ren:2019qow}
X.-L.~Ren, E.~Epelbaum and J.~Gegelia,
Phys. Rev. C \textbf{101}, 034001 (2020).

\bibitem{Baru:2019ndr}
V.~Baru, E.~Epelbaum, J.~Gegelia and X.-L.~Ren,
Phys. Lett. B \textbf{798}, 134987 (2019).

\bibitem{Bruns:2004tj}
P.~C.~Bruns and U.-G.~Mei{\ss}ner,
Eur. Phys. J. C \textbf{40}, 97-119 (2005).

\bibitem{Kubis:2000zd}
B.~Kubis and U.-G.~Mei{\ss}ner,
Nucl. Phys. A \textbf{679}, 698-734 (2001).

\bibitem{Bruns:2010sv}
P.~C.~Bruns, M.~Mai and U.-G.~Mei\ss ner,
Phys. Lett. B \textbf{697}, 254-259 (2011).

\bibitem{Morimatsu:2019wvk}
O.~Morimatsu and K.~Yamada,
Phys. Rev. C \textbf{100}, no.2, 025201 (2019).



\bibitem{Mai:2014xna}
M.~Mai and U.-G.~Mei\ss{}ner,
Eur. Phys. J. A \textbf{51}, 30 (2015).

\bibitem{Hemingway:1984pz}
R.~J.~Hemingway,
Nucl. Phys. B \textbf{253}, 742-752 (1985).

\bibitem{Doring:2011xc}
M.~D\"oring and U.-G.~Mei{\ss}ner,
Phys. Lett. B \textbf{704}, 663 (2011).

\bibitem{Amaryan:2020xhw}
M.~Amaryan \textit{et al.} [KLF collaboration],
[arXiv:2008.08215 [nucl-ex]].

\bibitem{Moriya:2013eb}
K.~Moriya \textit{et al.} [CLAS collaboration],
Phys. Rev. C \textbf{87}, no.3, 035206 (2013).


\bibitem{Miliucci:2021drx}
M.~Miliucci, A.~Amirkhani, A.~Baniahmad, M.~Bazzi, D.~Bosnar, M.~Bragadireanu, M.~Carminati, M.~Cargnelli, C.~Curceanu and A.~Clozza, \textit{et al.}
Acta Phys. Polon. Supp. \textbf{14}, 49 (2021).

\bibitem{Sakuma:2020btn}
F.~Sakuma \textit{et al.} [JPARC-E 015],
JPS Conf. Proc. \textbf{32}, 010088 (2020).





\end{references}
\end{document}